\documentclass[prl,twocolumn,superscriptaddress]{revtex4}
\usepackage{epsfig}
\usepackage{times}
\usepackage{amsmath}
\usepackage{amsfonts}
\usepackage{amssymb}
\usepackage[usenames]{color}
\usepackage{graphicx}

\newcommand{\beq}{\begin{equation}}
\newcommand{\eeq}{\end{equation}}
\newcommand{\beqa}{\begin{eqnarray}}
\newcommand{\eeqa}{\end{eqnarray}}

\newcommand{\NTT}{NTT Basic Research Laboratories, NTT Corporation, 3-1 Morinosato-Wakamiya, Atsugi, Kanagawa, 243-0198, Japan.}
\newcommand{\TQP}{NTT Theoretical Quantum Physics Center, NTT Corporation, 3-1 Morinosato-Wakamiya, Atsugi, Kanagawa 243-0198, Japan.}

\newcommand{\NII}{National Institute of Informatics, 2-1-2 Hitotsubashi, Chiyoda-ku, Tokyo 101-8430, Japan.}

\begin{document}
\title{Quantum metrology at the Heisenberg limit
\textcolor{black}{with the presence of}
independent dephasing
}

\author{Yuichiro Matsuzaki}
\email{matsuzaki.yuichiro@lab.ntt.co.jp} \affiliation{\NTT}\affiliation{\TQP}
 \author{Shiro Saito} 	\affiliation{\NTT}
 \author{William J. Munro}   \affiliation{\NTT}\affiliation{\TQP} \affiliation{\NII}
\begin{abstract}
The Heisenberg limit is the
 superior
 precision available by entanglement sensors.
 However, entanglement is fragile against dephasing, and
 there is no
 known
quantum metrology protocol that can achieve Heisenberg limited sensitivity with the presence of independent dephasing. 
  Here, we show that the Heisenberg limit is attainable under the effect
 of independent
dephasing
 under
 conditions where
 the probe qubits decohere due to
 both target fields and local environments.
 To detect the target fields,  
 we exploit the entanglement properties to decay much faster than the
 classical states due to collective noise
 while most of the previous schemes use a coherent phase
  shift from the target fields.
 Actually, 
 if the temporally fluctuating target fields
 behave as
 Markovian collective dephasing, 
  we can estimate the collective dephasing
  rate with a sensitivity at the Heisenberg
  limit under the effect
 of independent dephasing.
Our work opens the possibility for robust Heisenberg-limited
 metrology.
\end{abstract}

\maketitle
Quantum metrology is a \textcolor{black}{the field where one attempts to}
improve the performance of the \textcolor{black}{sensors measuring}
target fields
by using quantum properties
\cite{budker2007optical,balasubramanian2008nanoscaleetal,maze2008nanoscaleetal,toth2014quantum,degen2016quantum,degen2017quantum}.
Qubits \textcolor{black}{typically play the}
role as a probe to measure target fields
\textcolor{black}{when those}
qubits interact with the \textcolor{black}{fields we want to sense}.
When one prepares
\textcolor{black}{the qubits in a superposition state,}
\textcolor{black}{there are interference terms (non-diagonal elements) in the density matrix where }
the information of the target fields can be encoded.
Moreover, entanglement is considered a
resource to enhance this sensitivity \cite{wineland1992dj,
leibfried2004toward, giovannetti2004quantum,
giovannetti2006quantum,giovannetti2011advances,pirandola2017ultimate}.
If we use a separable state composed
of $L$ qubits to estimate
\textcolor{black}{our target fields with a parameter $\theta $,}
the uncertainty in this
estimation scales as $\delta \theta =O (L^{-1/2})$
\textcolor{black}{This is known as the standard quantum
limit (SQL).}
On the other
hand, it is in principle \textcolor{black}{possible without noise} to obtain a scaling of $\delta \theta =O
(L^{-1})$ by using  \textcolor{black}{an L-qubit entangled state}.
\cite{huelga1997improvement}.
\textcolor{black}{Such a scaling}
is
called the Heisenberg limit (HL) \cite{giovannetti2004quantum, giovannetti2006quantum,giovannetti2011advances}.

One of the \textcolor{black}{major} obstacles of quantum metrology is the fragility of the
\textcolor{black}{entangled probe state} against decoherence
\textcolor{black}{and especially dephasing}
\cite{huelga1997improvement,shaji2007qubit,demkowicz2012elusive}.
While
\textcolor{black}{these entangled probe states}
can be
strongly coupled with the target fields, those entangled states are
\textcolor{black}{highly}
sensitive
\textcolor{black}{to}
environmental noise.
It is crucial in the field of quantum metrology to construct a robust
entanglement sensor under the effect of realistic decoherence
\cite{chaves2013noisy,brask2015improved,smirne2016ultimate}.
\textcolor{black}{Actually,}
there are many proposals \textcolor{black}{to improve the sensitivity of the quantum sensors with realistic noise}
by using the quantum Zeno effect \cite{jones2009magnetic,matsuzaki2011magnetic,chin2012quantum,macieszczak2015zeno,tanaka2015proposed}, quantum error correction
\cite{kessler2014quantum,dur2014improved,arrad2014increasing,herrera2015quantum,
unden2016quantum,matsuzaki2017magnetic}, strong interaction between qubits \cite{dooley2018robust},
qubit motion
\cite{averin2016suppression,matsuzaki2016hybrid,matsuzaki2018quantum},
and even adaptive control \cite{pang2017optimal,demkowicz2017adaptive}.

Dephasing (or parallel noise) is considered a major challenge
\textcolor{black}{that needs to be overcome for}
the robust quantum metrology \cite{huelga1997improvement,lee2009optimization,escher2011general,nichols2016practical}.
Metrologically useful entanglement is typically designed to have a large
non-diagonal terms
where \textcolor{black}{such target field information is encoded.}
\textcolor{black}{Environmental} dephasing
parallel to the target
fields
induces a rapid decay of the
non-diagonal terms \textcolor{black}{where our target field information is encoded}. Such a decay significantly degrades the performance
of the quantum sensors. 
To recover the performance of the entanglement sensor,
there \textcolor{black}{is in principle a scheme}
to utilize the spatial correlation
\textcolor{black}{within} the environment that induces the dephasing
\cite{jeske2014quantum}.
\textcolor{black}{In such a case, one can}  achieve the 
\textcolor{black}{HL scaling only}
if the
form of the environmental spatial correlation satisfies \textcolor{black}{very} specific conditions
\cite{jeske2014quantum}.
\textcolor{black}{Currently however there} is no known metrological
protocol
\textcolor{black}{that achieves the HL scaling}
under the effect of independent dephasing where each local environment
independently couples with the probe qubits.
\textcolor{black}{It is generally thought that, under}
the independent Markovian dephasing,
an entanglement \textcolor{black}{based} sensor is metrologically equivalent
to the classical sensors
\textcolor{black}{as} the entanglement sensors cannot
beat the SQL \cite{huelga1997improvement,shaji2007qubit,demkowicz2012elusive}.
If the environment has a finite 
correlation time, the dephasing becomes non-Markovian,
and one achieves a sensitivity of $\delta \theta =O
(L^{-3/4})$, which beats the SQL but does not reach the HL
\cite{jones2009magnetic,matsuzaki2011magnetic,chin2012quantum,macieszczak2015zeno,tanaka2015proposed}. 

In this letter, we
\textcolor{black}{present} a sensing scheme
\textcolor{black}{that achieves}
the HL
under the effect of
 independent dephasing. 
  \textcolor{black}{Consider}
  that our $L$ probe qubits are
 affected by independent dephasing due to local environments,
 and \textcolor{black}{that we want to use}
 these probe qubits to measure a property of the target fields.
 \textcolor{black}{Pervious schemes typically wanted to measure the}
 amplitude of the time-independent target field using the probe qubits
  \cite{huelga1997improvement,shaji2007qubit, matsuzaki2011magnetic,chin2012quantum,macieszczak2015zeno,tanaka2015proposed}.
 On the other hand,
 \textcolor{black}{we can consider the situation where}
 the target fields are
 temporally fluctuating \textcolor{black}{and inducing}
 collective Markovian dephasing on those probe
 qubits. \textcolor{black}{In this situation our }
 purpose is to estimate the dephasing rate of this collective
 noise.
 We show that
 it is possible to estimate the collective dephasing
  rate  with \textcolor{black}{HL sensitivity}
  even under the effect of the independent dephasing.

\textcolor{black}{Let us}
describe our \textcolor{black}{scheme.}
Suppose that the target fields to
interact with the $L$ probe qubits  are
fluctuating \textcolor{black}{which induces decoherence.}
In this case, we can adopt a spin-boson model to describe the
interaction between the probe qubits and target fields
\cite{palma1996quantum,breuer2002theory}
\textcolor{black}{where}
each qubit
is affected by its local environment.
We define operators where $\hat{M}_z=\sum_{j=1}^{L}\hat{\sigma }_z^{(j)}$ denotes the
collective operator of the qubits, \textcolor{black}{$\hat{\sigma }_z^{(j)} =|1\rangle_j
\langle 1|-|0\rangle _j\langle 0|$} denotes the Pauli operator,  $\hat{b}_{k}$
($\hat{b}_{k}^{\dagger }$) denotes the annihilation (creation) operator of
the modes of the target fields, $\hat{c}_{j,k'}$ ($\hat{c}^{\dagger }_{j,k'}$) denotes the annihilation (creation) operator of
the local environmental modes coupled with a qubit at $j$-th site. We assume $[\hat{b}_k,
\hat{b}^{\dagger }_{k'}]=\delta _{k,k'}$, $[\hat{c}_{j,k'},
\hat{c}^{\dagger }_{j,k'}]=\delta _{j,j'}\delta _{k,k'}$.
The Hamiltonian is as following 
 \begin{eqnarray}
  H&=&H_S+H^{(\rm{ST})}_{\rm{I}}+H^{(\rm{SE})}_{\rm{I}}+H_{\rm{T}}+H_{\rm{E}} \nonumber \\
  H_S&=&\frac{\hbar \omega }{2}\hat{M}_z\nonumber \\
  H^{(\rm{ST})}_{\rm{I}}&=&\sum_{k}\hbar g_k\hat{M}_z
   (\hat{b}^{\dagger }_k +\hat{b}_k)\nonumber \\
    H^{(\rm{SE})}_{\rm{I}}&=&\sum_{j=1}^{L}\sum_{k'}\hbar \tilde{g}_{j,k'}\hat{\sigma }_z^{(j)}
   (\hat{c}^{\dagger }_{j,k'} +\hat{c}_{j,k'})\nonumber \\
  H_{\rm{T}}&=&\sum_{k}\hbar \omega _k\hat{b}^{\dagger
   }_k\hat{b}_k\nonumber \\
  H_{\rm{E}}&=&\sum_{j,k'}\hbar \omega '_{j,k'}\hat{c}^{\dagger }_{j,k'}\hat{c}_{j,k'} \nonumber \\
 \end{eqnarray}
where $\omega $ denotes the qubit \textcolor{black}{frequency},
$g_k$ denotes the interaction strength between
the qubits and the modes of the target, $\tilde{g}_{j,k'}$ denotes the interaction strength between
the qubit and the modes of the environment at $j$-th site, $\omega _k$ denotes the \textcolor{black}{frequency} of
the modes of the target fields, and $\omega '_{j,k'}$ denotes the \textcolor{black}{frequency} of
the modes of the environment at $j$-th site. 
\textcolor{black}{It is worth mentioning that, if a non-linear
interaction among qubits such as $H_I=\hbar g \hat{M}_z^2$ is available, a super
Heisenberg-limit is attainable to estimate the value of $g$
\cite{luis2004nonlinear,boixo2007generalized,roy2008exponentially,napolitano2011interaction,beau2017nonlinear}. However, here, we
consider a linear interaction $H^{(\rm{ST})}_{\rm{I}}$ where 
the HL is considered to be the ultimate precision \cite{giovannetti2004quantum, giovannetti2006quantum,giovannetti2011advances}.}
In the interaction picture, the Hamiltonian is
\begin{eqnarray}
&& H_I(t)=(\sum_{j=1}^{L}\hat{\sigma }_z^{(j)})\sum_{k}\hbar g_k
   (\hat{b}^{\dagger }_k e^{i\omega _kt}+\hat{b}_ke^{-i\omega
   _kt})\nonumber \\
   &&+\sum_{j=1}^{L}\sum_{k'}\hbar \tilde{g}_{j,k'}\hat{\sigma }_z^{(j)}
   (\hat{c}^{\dagger }_{j,k'}e^{i\omega '_{j,k'}t} +\hat{c}_{j,k'}e^{-i\omega '_{j,k'}t})
\end{eqnarray}
To characterize the property of the target fields (environment), we define a
power spectral density for the modes as
$J(\tilde{\omega} )=\sum_{k}\hbar ^2|g_k|^2 \delta (\tilde{\omega} -\omega
   _k)$ ($J'_j(\tilde{\omega} )=\sum_{k'}\hbar ^2|\tilde{g}_{j,k'}|^2 \delta (\tilde{\omega} -\omega
   _{j,k'})$).
   Although our main interest is to measure the collective dephasing rate
   with Markovian properties (that corresponds to a frequency-independent
   power spectral density), we adopt a more general setup of a Lorentzian  spectral
   density for the modes of the target fields (environment)
   such as $J(\tilde{\omega} )=\frac{1}{\pi }\frac{a/\tau _c}{(1/\tau _c)^2 +\tilde{\omega} ^2}$  
   ($J'_j(\tilde{\omega} )
    =\frac{1}{\pi }\frac{a'/\tau '_c}{(1/\tau '_c)^2 +\tilde{\omega} ^2}$)
   where $a$($a'$) denotes the amplitude and $\tau _c$ ($\tau '_c$) denotes the correlation
   time of the modes of the target fields (environment). It is worth
   mentioning that, by taking a limit of a small correlation time on the power
   spectral density, we can
   consider the Markovian behavior as a special case in this model. 
   We assume that
   the probe qubits, the target fields, and the local environments
   are separable at $t=0$.
     The initial state of the modes of target fields (environment) is a thermal
   equilibrium state such as $\rho
   _{\rm{T}}=\frac{1}{Z}e^{-\frac{H_{\rm{T}}}{k_BT}}$ ($\rho
   _{\rm{E}}=\frac{1}{Z'}e^{-\frac{H_{\rm{E}}}{k_BT}}$) where $T$ denotes
   the temperature, $k_B$ denotes the Boltzmann factor \textcolor{black}{respectively}. 
   $Z={\rm{Tr}}[e^{-\frac{H_{\rm{T}}}{k_BT}}]$ ($Z'={\rm{Tr}}[e^{-\frac{H_{\rm{E}}}{k_BT}}]$) denotes the
   renormalization factor. As an initial probe state, we \textcolor{black}{choose} the GHZ states
   $|\psi _{\rm{GHZ}}\rangle =\frac{1}{\sqrt{2}}(|00\cdots 0\rangle
   +|11\cdots 1\rangle )$.
   By tracing out the modes of the target
   fields and the environments with Born approximation,
   the decoherence dynamics of the probe
   qubits by
   the master equation in the Schrodinger picture is described by
\begin{eqnarray}
 && \frac{d\rho }{dt}
  =-i\omega [\hat{M}_z,\rho ]\nonumber \\
   &&-\sum_{k}|g_k|^2\frac{\sin \omega _kt}{\omega _k}
   \coth(\frac{\omega _k}{2k_{B}T})[\hat{M}_z,[\hat{M}_z, \rho
   ]]\nonumber \\
  &&-\sum_{j=1}^L \sum_{k'}|g_{j,k'}|^2\frac{\sin \omega _{k'}t}{\omega _{k'}}
   \coth(\frac{\omega _{k'}}{2k_{B}T}) [\hat{\sigma }^{(j)}_z,[\hat{\sigma
   }^{(j')}_z,\rho ]] \nonumber \\
   \label{master}
 \end{eqnarray}
  For the zero temperature of $T=0$, we can
 solve
 the master equation to \textcolor{black}{obtain}
      \begin{eqnarray}
    &&\rho (t)=\frac{1}{2}(|00\cdots 0\rangle \langle 00\cdots
     0|
    +|11\cdots 1\rangle \langle 11\cdots
     1|)
     \nonumber \\
     &+&\frac{e^{-i\omega t-L^2\Gamma
     _tt-L\gamma _tt}}{2}(|11\cdots 1\rangle \langle 00\cdots
     0|+|00\cdots 0\rangle \langle 11\cdots 1|)
     \nonumber 
   \end{eqnarray}
where $\Gamma _t=\frac{2a\tau _c^2}{t}(-1+e^{-\frac{t}{\tau _c}}+\frac{t}{\tau
_c})$ denotes the time-dependent collective dephasing rate and
$\gamma _t=\frac{2a'(\tau '_c)^2}{t}(-1+e^{-\frac{t}{\tau '_c}}+\frac{t}{\tau'
_c})$
denotes a time-dependent dephasing rate of the local environments \cite{palma1996quantum,breuer2002theory,chin2012quantum}. If the correlation time is much shorter than the typical time
of the dynamics which we call Markovian approximation,
 the dephasing
rate becomes time-independent. We define the Markovian dephasing rate
 of the target fields (environment)
as $\Gamma _{\rm{MC}}\equiv 2a\tau _c$ ($\gamma _{\rm{ME}}\equiv 2a'\tau '_c$). On the
other hand, in the limit of a long
target-fields (environmental)
correlation time, 
$\Gamma _t$ ($\gamma _t$) increases linearly against time.
 In this regime, we obtain $\Gamma _t\simeq at=\frac{\Gamma _{\rm{MC}}}{2\tau
_c}t$ ($\gamma _t\simeq a't=\frac{\gamma _{\rm{ME}}}{2\tau
'_c}t$). We define $\Gamma _{\rm{NMC}} \equiv
\sqrt{\frac{\Gamma _{\rm{MC}}}{2\tau_c}}$
($\gamma _{\rm{NME}}\equiv \sqrt{\frac{\gamma _{\rm{ME}}}{2\tau'_c}}$)
as a non-Markovian dephasing rate of the target fields.

We explain our protocol for the sensing by using
$L$ probe qubits for a given total time $T$. Assume that we can
prepare and readout the probe qubits with a time scale much faster than the
coherence time of the probe qubits.
First, we prepare the GHZ state of the $L$ probe
qubits. \textcolor{black}{Second, we let} the probe qubits evolves for a time $t$ according
to the master equation in the Eq. (\ref{master}).
\textcolor{black}{Third, we then} perform a measurement with a projective operator of
$\hat{\mathcal{P}}=|\psi _{\rm{read}} \rangle \langle \psi _{\rm{read}}|$.
 Finally, we repeat these process
$N\simeq T/t$ times.
The uncertainty to estimate a parameter $\theta $ of the target
is described as $\delta \theta =\frac{\sqrt{P(1-P)}}{|\frac{dP}{d\theta }|}\frac{1}{\sqrt{N}}$
\cite{huelga1997improvement}
where $P={\rm{Tr}}[\rho (t)\hat{\mathcal{P}}]$ denotes a probability
distribution and $\rho (t)$ denotes a density matrix of the probe qubits
at a time $t$.
Since our model is
general, our results include previously studied schemes
\cite{huelga1997improvement,matsuzaki2011magnetic,chin2012quantum,tanaka2015proposed} as special cases.
   \begin{table*}
\begin{center}\textcolor{black}{
    \begin{tabular}{| l | l | l | l | p{4.8cm} |}
    \hline
     &Markovian independent dephasing environment & Non-Markovian
     independent dephasing environment
     \\ \hline
      Markovian collective dephasing fields & \scalebox{1.1}{\ \ \ \ \ \ \ \ \
         \ $\delta \Gamma _{\rm{MC}}=O(L^{-1})$}
         &\scalebox{1.1}{\ \ \ \ \ \ \ \ \
         \ $\delta \Gamma _{\rm{MC}}=O(L^{-1})$} 
                     \\ \hline
     Non-Markovian collective dephasing fields&
         \scalebox{1.1}{\ \ \ \ \ \ \ 
         \ $\delta \Gamma _{\rm{NMC}}=O(L^{-1/2})$ } &
              \scalebox{1.1}{\ \ \ \ \ \ \ 
         \ $\delta \Gamma _{\rm{NMC}}=O(L^{-1/2})$ }
     \\ \hline
    \end{tabular}}
 \caption{
 Performance of our sensing scheme where $L$ probe qubits
 interacts with both target fields and local environments.  The target
 fields are temporally fluctuating
 \textcolor{black}{which} induces collective dephasing
 on the probe qubits.
  Surprisingly, under the effect of independent dephasing due to the
 local environments, we can achieve a Heisenberg limit \textcolor{black}{scaling} when the target fields have a
 Markovian (or time local) \textcolor{black}{nature}.
 On the other hand, if the target fields have a
 memory effect, the property becomes non-Markovian (or time non-local), and we cannot even beat
 the standard quantum limit under the effect of independent dephasing.
 }
 \label{table}
\end{center}
   \end{table*}

Let us review the previous  quantum metrology to measure the
amplitude of time-independent target fields \cite{huelga1997improvement,matsuzaki2011magnetic,chin2012quantum,tanaka2015proposed}. We assume the
amplitude of the target fields has a linear relationship with the frequency
$\omega $ and this amplitude is weak.
The aim in these research is to estimate the value of $\omega $.
Also, in these calculations \cite{huelga1997improvement,matsuzaki2011magnetic,chin2012quantum,tanaka2015proposed},
the collective dephasing is not considered, and so we set $a=0$. The
uncertainty of the estimation is given as $\delta \omega =\frac{{\rm{Exp}}[L\gamma _tt]}{L\sqrt{Tt}}$
where we choose $|\psi _{\rm{read}}\rangle =\frac{1}{\sqrt{2}}(|0\cdots
0\rangle +i|1\cdots 1\rangle )$.
For the
independent Markovian environment
with
a short $\tau
_c'$, we obtain $\delta \omega =\frac{{\rm{Exp}}[L\gamma
_{\rm{ME}}t]}{L\sqrt{Tt}}$, and this scales as $\delta \omega
=O(L^{-1/2})$ by taking an optimized interaction time  as
$t=O(L^{-1})$ \cite{huelga1997improvement}.
On the other hand,
for the
independent non-Markovian environment
with
a long $\tau
_c'$, we obtain $\delta \omega
=\frac{{{\rm{Exp}}}[La't^2]}{L\sqrt{Tt}}$,
\textcolor{black}{which}
scales as $\delta \omega
=O(L^{-3/4})$ by taking for an optimized interaction time as
$t=O(L^{-1/2})$
\cite{jones2009magnetic,matsuzaki2011magnetic,chin2012quantum,tanaka2015proposed}.
To estimate the amplitude of the time-independent
target fields, the non-Markovian properties of dephasing contribute to improve
the sensitivity of the entanglement sensor.
However, in either case, we cannot achieve the HL
under the effect
of the independent dephasing.


We \textcolor{black}{can} show that, for the estimation of the Markovian collective dephasing rate due to the
temporally fluctuating target fields, we can achieve the
HL
under the effect of independent Markovian dephasing.
More specifically, we \textcolor{black}{can} calculate the uncertainty of the estimation of
with a white noise  power spectral density $J(\tilde{\omega} )=\frac{a\tau _c}{\pi
}$ where we take a limit of a small correlation time for the Lorentzian
power spectral density. 
 Since we assume that the qubit frequency $\omega $ is known for this
 estimation, we can ignore this effect.
 \textcolor{black}{Now,}
 let us discuss the case of using a separable state of
 the $L$ probe qubits for the estimation of $\Gamma _{\rm{MC}}$.
For a single qubit sensor with an initial state of $|+\rangle
 =\frac{1}{\sqrt{2}}(|0\rangle +|1\rangle )$, we obtain
$ \delta \Gamma _{\rm{MC}}=\frac{\sqrt{1- {\rm{Exp}}[-2\gamma _tt-2\Gamma_{\rm{MC}}t
 ]}}{\sqrt{Tt}{\rm{Exp}}[-\gamma _tt-\Gamma_{\rm{MC}}t ]}=O(L^0)$.
 By using $L$ qubits in
 parallel as a separable state,
 the sensitivity can be enhanced by a factor of $\sqrt{L}$ due to a
 central limit theorem, and so the uncertainty of the separable sensor
 is $\delta \Gamma
 _{\rm{MC}}=O(L^{-1/2})$, which is bounded by the SQL.
Next, we \textcolor{black}{can} calculate the uncertainty 
with the GHZ states composed of $L$ probe qubits as
\begin{eqnarray}
 \delta \Gamma _{\rm{MC}}=\frac{\sqrt{1- {\rm{Exp}}[-2L\gamma _tt-2L^2\Gamma_{\rm{MC}}t
 ]}}{L^2\sqrt{Tt}{\rm{Exp}}[-L\gamma _tt-L^2\Gamma_{\rm{MC}}t ]}.\label{mcrate}
\end{eqnarray}
for $|\psi _{\rm{read}}\rangle =\frac{1}{\sqrt{2}}(|0\cdots
0\rangle +|1\cdots 1\rangle )$.
  By choosing $t=t_0 /L^s$
 where $t_0$ denotes a constant time and $s$ denotes a constant value,
 we obtain $\delta \Gamma
 _{\rm{MC}}=\frac{\sqrt{1- {\rm{Exp}}[-2L^{1-s}\gamma _tt_0-2L^{2-s}\Gamma_{\rm{MC}}t_0
 ]}}{L^{2-s/2}\sqrt{Tt_0}{\rm{Exp}}[-L^{1-s}\gamma _tt_0-L^{2-s}\Gamma_{\rm{MC}}t_0 ]}$.
 The uncertainty becomes $\delta \Gamma
 _{\rm{MC}}=\frac{\sqrt{1- {\rm{Exp}}[-2\gamma _tt_0/L-2\Gamma_{\rm{MC}}t_0
 ]}}{L\sqrt{Tt_0}{\rm{Exp}}[-\gamma _tt_0/L-\Gamma_{\rm{MC}}t_0 ]}$ for $s=2$.
For a large $L$ the effect of the independent 
 dephasing becomes negligible regardless of the correlation time of the environment, and the uncertainty is approximated as $\delta \Gamma
 _{\rm{MC}}\simeq \frac{\sqrt{1- {\rm{Exp}}[-2\Gamma_{\rm{MC}}t_0
 ]}}{L\sqrt{Tt_0}{\rm{Exp}}[-\Gamma_{\rm{MC}}t_0 ]}=O(L^{-1})$. 
 Therefore, we achieve the
 HL
 under the effect of independent dephasing.

 We explain intuitive reasons why we can achieve the HL
 to estimate the collective Markovian dephasing rate by using the entanglement. It is worth
 mentioning that, if the initial state of the probe qubit is the GHZ
 state,
 the collective Markovian dephasing occurs in a time scale of
 $t=O(L^{-2})$, while independent Markovian (non-Markovian) dephasing
 occurs in a time scale of  $t=O(L^{-1})$ ($t=O(L^{-1/2})$). 
 This means that we can observe the change in the dynamics
 of the probe qubits due to the collective decay within a time scale of
 $t=O(L^{-2})$ while the effect of the independent dephasing
 is negligible within this time scale for a large $L$. Moreover, since
 it takes a time of  $t=O(L^{-2})$ for a single measurement, we can
 repeat the measurements $N\simeq T/t=O(L^2)$ times for a given time
 $T$. Therefore, we can decrease the uncertainty of the estimation of
 the collective dephasing rate by $\delta \Gamma _{\rm{MC}}=O(N^{-1/2})=O(L^{-1})$,
 which achieves the HL.

\textcolor{black}{Now for comparison,}
we calculate the uncertainty to estimate non-Markovian collective
dephasing rate
$\Gamma_{\rm{NMC}}$ under the effect of independent dephasing.
Here, we take the limit of a long correlation time $\tau _c$ for the
target fields. The noise  power spectral
density is described as $J(\tilde{\omega })=  a\delta (\tilde{\omega } )$.
Similar to the Markovian case,
the uncertainty to estimate
 $\Gamma_{\rm{NMC}}$ is bounded by the SQL
 if we use $L$ probe qubits as a separable
 state.
On the other hand,
with an entanglement,
we obtain
\begin{eqnarray}
 \delta \Gamma _{\rm{NMC}}=\frac{\sqrt{1- {\rm{Exp}}[-2L\gamma _tt-2L^2\Gamma_{\rm{NMC}}^2t^2
 ]}}{2L^2\Gamma _{\rm{NMC}}t\sqrt{Tt}{\rm{Exp}}[-L\gamma _tt-L^2\Gamma _{\rm{NMC}}^2t^2 ]}
\end{eqnarray}
  By choosing $t=t_0 /L^s$, we obtain $\delta \Gamma _{\rm{NMC}}=\frac{\sqrt{1- {\rm{Exp}}[-2L^{1-s}\gamma _tt_0-2L^{2-2s}\Gamma_{\rm{NMC}}^2t^2_0
 ]}}{2L^{2-3s/2}\Gamma
 _{\rm{NMC}}t_0\sqrt{Tt_0}{\rm{Exp}}[-L^{1-s}\gamma _tt_0-L^{2-2s}\Gamma
 _{\rm{NMC}}^2t_0^2 ]}$.
 This uncertainty \textcolor{black}{is}
 minimized
 \textcolor{black}{when}
 $s=1$ such
 \textcolor{black}{that}
 $\delta \Gamma _{\rm{NMC}}=\frac{\sqrt{1- {\rm{Exp}}[-2\gamma _tt_0-2\Gamma_{\rm{NMC}}^2t^2_0
 ]}}{2L^{1/2}\Gamma
 _{\rm{NMC}}t_0\sqrt{Tt_0}{\rm{Exp}}[-\gamma _tt_0-\Gamma
 _{\rm{NMC}}^2t_0^2 ]}=O(L^{-1/2})$, which is the SQL.
 Therefore, to estimate the non-Markovian
 collective dephasing rate, the entanglement sensor does not offers a
 scaling advantage over the separable sensor.

 We explain the reason why we cannot beat the SQL to estimate the
 non-Markovian collective dephasing rate.
 Non-Markovian dephasing occurs in a time scale of $t=O(L^{-1})$.
 This means that
 it takes a time of  $t=O(L^{-1})$ for a single measurement, we can
 repeat the measurements $N=T/t=O(L)$ times for a given time
 $T$. So the uncertainty of the estimation of
 the non-Markovian collective dephasing rate is given $\delta \Gamma
 _{\rm{MC}}=O(N^{-1/2})=O(L^{-1/2})$, which is the SQL.
 
 Our results (summarized in \textcolor{black}{Table} \ref{table}.)
 are essentially different
 from the previously studied cases of measuring the amplitude of
 the time-independent fields under the effect of independent dephasing
 \cite{huelga1997improvement,matsuzaki2011magnetic,chin2012quantum,tanaka2015proposed}.
 In the previous cases, non-Markovian properties of the local
 environment let us beat the SQL
 \cite{matsuzaki2011magnetic,chin2012quantum,tanaka2015proposed}, while
a Markovian environment
\textcolor{black}{made} the entanglement sensor metrologically
 equivalent to the separable
 \textcolor{black}{ones}
 \cite{huelga1997improvement}.
 \textcolor{black}{Non}-Markovian properties
 \textcolor{black}{were}
 important to beat the SQL.
 \textcolor{black}{On the other hand,}
 Markovian properties
 of the target fluctuating fields actually helps to achieve the HL in
 our case,
 while non-Markovian properties of the target fluctuating fields
 destroy the
 advantage of the entanglement sensor. 

  \begin{figure}[h!] 
\begin{center}
\includegraphics[width=0.85\linewidth]{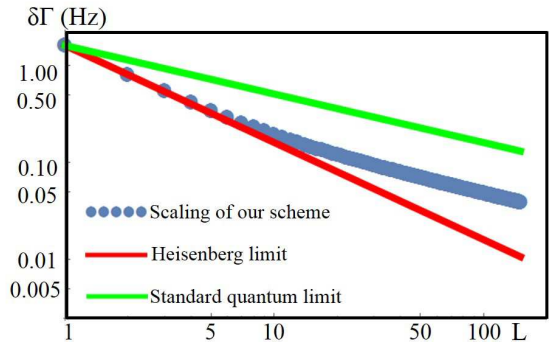} 
\caption{
 \textcolor{black}{Plot of}
 the uncertainty $\delta \Gamma _{\rm{MC}}$ (blue dots)
 against the
 number of the probe qubits 
 where we estimate
 the collective dephasing rate due to the temporally fluctuating target fields
  under the effect of
 independent Markovian dephasing due to local environments. 
 The parameters are set as $\Gamma _{\rm{MC}}=1$ (Hz),
 $\gamma _{\rm{ME}}=0.2$ (Hz), $T=1$ (s) and
 $\tau _c=0.001$ (s). The red (green) line shows the scaling of the Heisenberg limit (SQL).
 For a small number of qubits, the time scale of the collective
 dephasing is shorter than the correlation time $\tau _c$, and so the
 fluctuating target fields can be approximately treated as collective Markovian dephasing. In
 this regime, we can estimate $\Gamma _{\rm{MC}}$ with a sensitivity of
 the Heisenberg limit.
 On
 the other hand, for a large number of the qubits, the time scale of the
 collective dephasing becomes shorter than the correlation time $\tau _c$, and the
 collective dephasing shows a non-Markovian property. In this regime,
 the sensitivity is bounded by the SQL.
 }
 \label{scale}
\end{center}
\end{figure}

Let us now calculate the uncertainty of the estimation
\textcolor{black}{when we have} a finite
correlation time $\tau _c$ for the target fields. 
While we \textcolor{black}{can} analytically calculate the uncertainty of the estimation
\textcolor{black}{in the limits of}
short or a
long correlation times,
\textcolor{black}{will consider the finite $\tau _c$ situation now,}
and so we numerically plot the uncertainty of the estimation of the
collective dephasing rate in the Fig. \ref{scale}.
Here, we choose the interaction time $t$ to minimize the uncertainty,
and
assume that the local environment
\textcolor{black}{is Markovian.}
We observe a clear transition of the scaling from the
HL
to the SQL as we increase the number of the probe
qubits. This can be understood as follows. For a small number of the
qubits, the characteristic time of the collective dephasing is much
longer than the correlation time, and so we can use the Markovian
assumption.
On the other hand, as we increase the number of the qubits, the
collective dephasing becomes stronger, and the characteristic
time of the collective dephasing will be ultimately shorter than the
correlation time. This means that, in the limit of a large $L$,
the target fields should show the non-Markovian properties.
 From the \textcolor{black}{Table} \ref{table},
such a change of the property of the target fields clearly affects
the uncertainty of the estimation, which induces the transition of the
scaling from the HL
to the SQL.
It is worth mentioning that, although we
cannot achieve the
HL
for a large $L$ with a finite
correlation time $\tau _c$, we can still
obtain a constant factor improvement with the entanglement sensor over the classical sensors, as
shown in the Fig. \ref{scale}.

In conclusion, we \textcolor{black}{have shown} that
the 
Heisenberg limit is attainable in
quantum metrology under the effect of independent dephasing.
We \textcolor{black}{consider the situation where the}
probe qubits interacts with both the target fields and
local environments. \textcolor{black}{More importantly we were interested in the situation where the}
target fields are temporally fluctuating
\textcolor{black}{which induces}
 Markovian collective dephasing, while the local environment
 \textcolor{black}{only} induces
 independent dephasing. We find that,
 \textcolor{black}{when estimating}
 the collective
 dephasing rate due to the target fields, we can achieve the Heisenberg
 limited \textcolor{black}{scaling} with an entanglement sensor.
 \textcolor{black}{This in turn paves the way for a future generation of
 HL sensor measuring fluctuating field.}
\textcolor{black}{Moreover, our results are essential to understand the
ultimate limit of the entanglement sensor with realistic conditions.}


This work was supported
in part by CREST (JPMJCR1774), JST and
the MEXT Grants-in-Aid for Scientific Research on Innovative Areas ''Science of
Hybrid Quantum Systems''(Grant No. 15H05870).


\end{document}